# Multimodality and parallelism in design interaction: co-designers' alignment and coalitions

Françoise DETIENNE[a,1] and Willemien VISSER[a]

[a] *EIFFEL2 (Cognition & Cooperation in Design) - INRIA (National Institute for Research in Computer Science and Control), France*

**Abstract**. This paper presents an analysis of various forms of articulation between graphico-gestural and verbal modalities in parallel interactions between designers in a collaborative design situation. Based on our methodological framework, we illustrate several forms of multimodal articulations, that is, integrated and non-integrated, through extracts from a corpus on an architectural design meeting. These modes reveal alignment or disalignment between designers, with respect to the focus of their activities. They also show different forms of coalition.

## Introduction

Collaborative design takes place through interaction between designers. We will see that this apparently unequivocal statement, which may even seem tautological, conveys characteristics of design thinking that are essential, in our view. During co-design sessions, individual design plays of course —also— an important role [see e.g., 1, 2]. Yet, an essential part of collaborative design thus takes place —that is progresses— through interaction. This interaction takes different forms and refers to various representation-construction activities [see 3]. The different forms that interaction may take in collaborative design —verbal, graphical, gestural, postural— are, in our view, not the simple expression and transmission of ideas previously developed in an internal medium, developed in a kind of *mentalese* (such as Fodor's "language of thought"). They are more, and of a different nature, than the trace of a so-called "genuine" design activity, which would be individual and occur internally, and which verbal and other forms of interaction would allow to share with colleagues, once it has been developed.

In this paper, we present an analysis of various forms of articulation between graphico-gestural and verbal modalities in interactions between designers in a collaborative design situation. After a brief introduction of the theoretical framework that we adopt for design interactions, we present our methodological framework and illustrate several forms of multimodal articulations, that is, integrated and non-integrated, through extracts from a corpus on an architectural design meeting. This corpus was collected in the context of the MOSAIC project conducted in the COGNITIQUE program "Cognitions, interactions sociales, modélisations" [4]. The meeting, which took place in a preliminary phase of a renovation project, involved three architectural designers, Charles, Louis, and Marie.

---

[1] Corresponding author: Françoise Détienne, EIFFEL2, INRIA, bât. 23, Rocquencourt B.P. 105, 78153 Le Chesnay Cedex, France

## 1. From individual to collaborative design

Since the early days of research on design activities [e.g., 5], many, if not most, empirical design studies, especially the cognitive ones, use simultaneous verbalization [6, 7]. Asking people to "verbalize their thoughts" or to "think aloud" is, however, only necessary for data collection on individually conducted activities. People working together do "naturally" express their thoughts —or, at least, part of them. The analysis of the two families of corresponding corpora requires different specific methods.

Our position is that going from individual to collaborative design merits emphasis on two new foci: *multimodality*, referring to the importance of the graphico-gestural dimensions in relation to the verbal dimension of interaction, and *parallelism*, referring to the importance of activities that one or several co-designers conduct in parallel (simultaneously, or with more or less overlapping) in addition to those they conduct in sequence.

"Multimodality" refers to the use of various semiotic systems (verbal, graphical, gestural, postural).

Our use of the term "parallelism" covers both strictly simultaneous actions, and actions with more or less overlapping between them.

### 1.1. From merely verbal to multimodal interactions

Many previous studies of design, for example on software design, have analysed collaborative activities that take place in face-to-face meetings, such as idea-generation and technical review activities [8-11]. In these studies, researchers have identified various types of collaborative design activities based on verbal interactions between co-designers.

One set of collaborative activities is related to *activities on the objects of design*, the artefacts. These activities concern the evolution of the design problem and solution, for example, elaboration of the problem, generation of a solution and identification or enhancement of alternative solutions. Are also of this kind evaluative activities —for example, the evaluations of solutions or alternative solutions based on argumentation.

A second type of activity concerns the *construction, by a group of co-designers, of "common reference", or "common ground"* [what Visser prefers to qualify as "inter-designer compatible representations", 3]. For example, clarification or cognitive synchronization activities take place when a group negotiates or constructs such common representations of the current state of the solution.

*Group management activities* are a third kind of design activity. These activities are frequently related to issues of process. Project-management activities that concern the coordination of people and resources —the allocation and planning of tasks— are of this kind. Meeting-management activities —the ordering and postponing of discussion topics— are another example.

All these activities, which characterize collaborative design, do not only occur in a verbal mode, but also in graphical and gestural modes. The importance of graphical representations, as intermediate or intermediary representations, has been underlined in the literature on design. However, there have been only few attempts to systematize the analysis of these various modes of interaction in design, especially their construction rather than their use [3]. This is the line of research adopted in this paper.

## 1.2. From sequential to parallel interactions

Based on the analysis of verbal interactions, a body of work has focused on the types of activities occurring in design meetings and their sequential organization [9, 11-13], such as sequences of "moves" or "turns" in the argumentation process. Accounting for designers' spontaneous sequential organization of activities is of particular interest with respect to design methods that specify steps in design.

D'Astous, Détienne, Visser and Robillard [9], for example, analysed the argumentative moves in software technical review meetings. One of their results was that the elaboration of a solution tends to be followed by either its evaluation or the development of an alternative solution. In the second case, there is an implicit negative evaluation of the previously proposed solution.

In another study, Détienne, Martin and Lavigne [12] examined the negotiation patterns leading participants to converge in multidisciplinary meetings in aeronautical design. They found a typical temporal negotiation pattern composed of three steps: (1) analytical assessment of the current solution, that is, systematic assessment according to constraints; (2) if step 1 did not lead to a consensus, comparative or/and analogical assessment; (3) if step 2 did not lead to a consensus, use of one or several argument(s) from authority.

With regard to our particular interest in multimodal interaction, we shift our focus from a view that analyses sequences of actions, to a view that analyses how strictly simultaneous or more or less overlapping actions are articulated. Focus is then no longer on the sequential organization of activities (which is still quite relevant), but rather on the articulation of activities that one or more designers implement in parallel (that is, in strict simultaneity or with more or less overlapping between the activities).

Accounting for parallel activities is particularly relevant for analysing the alignments and oppositions between designers during their collaborative activity. It is also of special interest for the development of computer tools to support collaborative design, such as augmented-reality environments that enable synchronous collaboration without imposing a master/slave mode.

## 2. Taking into account graphics and gestures in collaborative design

Intermediate representations are the representations that clearly occupy the greatest part of the design activity during a project. Graphical and gestural interactions play a role that is, at least, as important in the construction of these representations as purely verbal expressions.

### 2.1. Intermediate and intermediary representations

The representations produced and used in early and later intermediate design phases are generally not of the same type as the final representations, which specify the implementation of the artefact. They allow designers to focus on different aspects of their design [14], which may or may not be maintained until the final design stages.

In addition to being intermediate between the requirements at the start of a design project and the specifications at its other extremity, representations have also an intermediary function. Two types of intermediary representations are to be distinguished: they can be intermediary between designers and their artefact, and between several designers. In their first role, they function as tools and are often qualified as "cognitive artefacts", by reference to Norman [15]. With respect to their second role, Boujut and Laureillard [16] or Schmidt and Wagner [17] propose the concepts of "cooperative features", "coordinative artefacts" and "intermediary objects" to characterize the particular role that these intermediary representations play in collaborative processes. They may have functions such as construction of common ground concerning design principles, or tasks; reminders of such principles, and open problems; traces of activities; and representations of design decisions. In this way, they may support co-design, argumentation, explanation, and simulation, or be an external memory of design rationale [17].

## 2.2. Graphico-gestural representations

In semiotics, ethology, and more recently pragmatic linguistics and psycholinguistics, analysis of gestural interaction represents already a considerable body of research work [18-21]. Often referring to "workplace studies", ethnography or ethnomethodology for their theoretical and methodological position [24, 25], many authors nowadays mainly present their data and results in narrative, anecdotal terms, providing rather detailed descriptions of "cases", but without much generalisation (or generalisability) in their results and conclusions. For instance, Brassac and Le Ber [22, see also 23] present detailed descriptions of co-present agronomists and computer scientists collectively designing a knowledge-based system, using "cognitive, corporal, documentary and material" resources. The authors describe verbal (oral), gestural, and graphical (both writing and drawing) activities, showing several examples of interaction between verbal and graphical activities.

In the cognitive ergonomics of design, research on graphico-gestural interaction is at its beginnings [26, 27]. An important difference with more narratively oriented approaches is our aim to reach generalisable results concerning design and to be able to compare different design situations with respect to explicit dimensions. Up to date, cognitive ergonomics has examined graphical and other types of external representations, but mostly the representational structures, not their elaboration [3].

In an empirical study on collaborative design in a technology-mediated situation, Détienne, Hohmann and Boujut [28] showed that graphical representations of the design artefact played a central role. In the synchronous mode, whiteboard and shared CAD applications were used to co-produce solutions and to support argumentation and explanation. Supporting online co-production activity was the most frequent use of technical devices. Computer graphics and sketches on the Netmeeting whiteboard supported this activity.

It is not only in distant interaction, however, that other than verbal representations play an important role. In their analysis of small group conceptual design sessions in co-presence, Tang [29] and Tang and Leifer [30] have identified the importance of gesture, in addition to graphical representations. They have proposed a framework for the analysis of workspace activity that establishes relationships between actions that occur in the workspace and their functions. The "conventional view" of workspace activity considers this space as "primarily a medium for storing information and conveying ideas through listing text and drawing graphics". The authors aim to extend this conventional view, adding three other aspects to workspace activity: "gestural expression", "developing ideas", and "mediating interaction" [30, p. 247]. Besides drawing or sketching, already identified in previous research to occur often in collective-design meetings, gesture was found to take place frequently. The main function of gesture was to mediate interaction between the different design participants: more than half of the gestures fulfilled this function through participants engaging or asking for attention.

On the website page that presents the research on gesture in her STAR team (Space, Time, and Action Research, retrieved November 24, 2005, from http://www-psych.stanford.edu/~bt/gesture/), Tversky notices that "although it is typically thought that gestures accompany speech, gestures often accompany listening (Heiser, Tversky, MacLeod, Carletta, and Lee, in preparation) and non-communicative thinking (Kessell, 2004). In both cases, they seem to serve to augment spatial working memory, much as sketching a diagram would." Tversky also refers to research in which the combined use of graphics and gesture, rather than verbal expression, was identified. "In collaboration with diagrams, dyads save speech by pointing and tracing on the diagram. Partners look at the diagrams and their hands, not at each other (Heiser, Tversky, and Silverman, 2004). Having a shared diagram to gesture on facilitates establishing common ground and finding a solution. It also augments solution accuracy."

## 3. Our method: articulating modalities

In order to analyse the articulation between modalities in a collaborative design setting, we have adopted a functional perspective based on local design goals that interlocutors may share or not, at a particular moment in their interaction. Our distinction between local goals is based on the pursuit of the functional design activities identified previously in our COMET method [31].

### 3.1. The COMET method

With our colleagues of the CNAM [see e.g., 32], we have developed COMET for the analysis of collaborative design processes [31], integrating protocol analysis as developed for the analysis of individually conducted activities, and pragmatic linguistics' verbal-interaction analysis [33][2].

Underlying our development and use of this method, is our aim to formulate a generic model of the socio-cognitive aspects of collaborative designing. The descriptors (categories) distinguished in COMET are design actions and objects that numerous empirical cognitive design studies have shown to be characteristic of designing.

According to COMET, verbal turns are cut up into one or more individual Units according to a coding scheme developed on a Predicate(Argument(s)) basis. Each predicate only admits a number of possible arguments. Predicates (ACT) correspond to actions implemented by participants; arguments (OBJ) correspond to objects concerned by the action (the actor, the object of the action, tools and other elements involved; see the examples presented hereafter). According to the form of the predicate (Assertion or Request), each unit is modulated (MOD). The default value of a unit is assertive: modulation is coded explicitly only if its predicate is a request. Thus, each Unit is coded as MOD[ACT/OBJ], where MOD may be absent —in which case it is assertive (see Table 1).

---

[2] In COMET, we did not introduce, however, the means to analyse linguistic phenomena such as modalisation procedures, for example expressions of addressing or of politeness [Araújo Carreira, 2005 #2626] Neither did we introduce the means to describe graphical and gestural interaction.

**Table 1.** Basic coding scheme, presenting the elements of each category [from 31]

| Modulation (MOD) | Predicate (ACT) | Argument (OBJ) |
|---|---|---|
| Assertion | Generate    (GEN)<br>Proposing a new element into the dialogue (a solution, a goal, an inferred data, etc.) | Problem data (DAT) |
| Request (REQ) | Evaluate    (EVAL)<br>Judging the value of a subject. This evaluation can be negative, positive, or neutral. | Solution elements (SOL) |
| | Inform    (INFO)<br>Handing out new knowledge with respect to the nature of a subject | Domain objects (OBJ) |
| | Interpret    (INT)<br>Expressing a personal representation of a subject. This representation is made by expressions such as "I believe that…", "I think …" or "…maybe…". | Domain rules or procedures (PROC) |
| | | Goal (GOAL) |
| | | Task (TASK) |

In this paper, we introduce a modification in the SOL category in order to distinguish between solutions, depending on their reference to a problem. We consider that solutions are associated with problems (which constitute a kind of superordinate category with respect to the other arguments). Two solutions SOL1_PBp and SOL2_PBp to the problem PBp belong to the same category, whereas two solutions to different problems, SOL3_PBq and SOL4_PBr, belong *per se* to different object categories.

Using these predicate and argument categories, we establish two distinctions between activities, one depending on their type of predicate, and the other depending on their argument. The first distinction differentiates elaborative (Generate), evaluative (Evaluate) and clarification activities (Inform and Interpret). The second distinction tells apart three groups of activities: activities in the group space (Task or Goal), the problem/solution space (Problem data and Solution elements attached to a same problem), and the domain space (Domain objects, Domain rules or procedures).

### 3.2. Description of graphico-gestural activities

The endeavour in which we are engaged at present consists in extending the analysis of verbal interactional data to that of other semiotic systems, that is, to analyse design interaction's multimodality. Up to now, we have developed a description language for the graphico-gestural activities [26, 27], and we are examining the articulation between graphico-gestural and verbal dimensions in collaborative design interaction. This has been applied to our corpus of architectural design and specifically to the analysis of the overhead view (top right view in Figure 1).

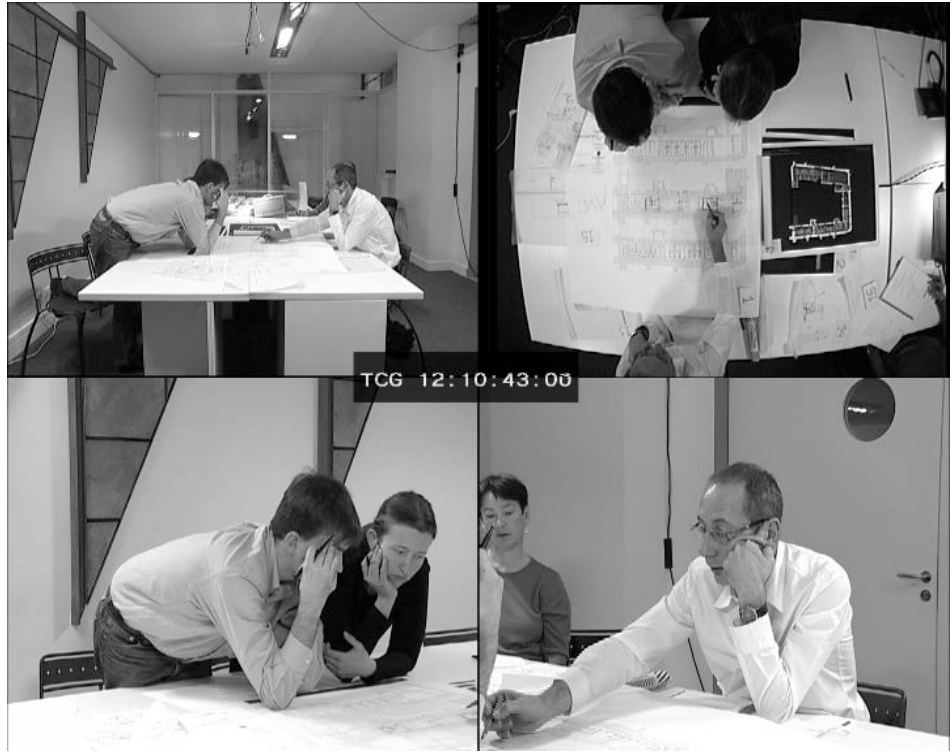

**Figure 1.** The four views of the MOSAIC corpus

Our graphico-gestural description language accounts for movements performed, by hand with or without a tool (pen, pencil, ruler or other), on external representations, in particular, plans and draft papers. It uses the same predicate structure as COMET, i.e., Predicate(Argument(s)), which corresponds to Action(Object(s)) units, completed by their duration and localisation. Each graphico-gestural unit is the description of one action performed by a designer, from time t0 to time t1, on an object (plan, draft paper, or other document), in a particular area of the table, by hand only or with a tool. The graphico-gestural actions are: Point (pointing), Delimit_2D (delimitation in two dimensions), Delimit_3D, Graph_trac (graphical tracing or drawing), Text_trac (textual tracing or writing), Moving (e.g. moving a plan from a peripheral to a central area), Rotating (e.g. rotating a plan in someone's direction), and Overlaying (e.g. overlaying a plan with a draft paper).

*3.3. Integration between activities*

Based on the COMET frame, enlarged in order to account for multimodal interaction, designers are considered to pursue the same local goal if their objects are in the same category (at a more or less global level), even though the modalities adopted to pursue them are different. In this case, we consider that their *activities* are *integrated* [27].

Different designers working in an integrated mode may thus perform each one the same type of action on what constitutes, at a more or less global level, the same object (e.g., generate together a solution to a particular problem, or generate alternative solutions to this problem), or perform different types of actions on the same object (e.g., one generates, the other evaluates a solution). In both cases, there is an *alignment between the designers* with respect to their sharing a same focus of work, and, more generally, a same local goal. The actions may be conducted through identical or different modalities, which may be redundant or complementary [35, 36]. In this text, we focus on examples of activities that are performed through different modalities.

For example, two designers working on the same object (e.g., the same alternative solution or two alternative solutions to a particular problem) may conduct the same action (generation) through two modalities, verbal and graphical: one designer verbally formulates the solution, while a colleague draws or details it in a sketch (i.e., works on the same global solution, even if she develops a different elementary solution). Two designers working on the same object may also conduct different actions on it through distinct modalities. One designer generates a solution in a verbal mode, while the other evaluates the solution by simulating it through gestures.

Designers are considered to pursue different local goals if the objects on which they are working come from different categories (e.g. solutions to distinct problems or a solution vs. a task). In this case, we consider that their *activities* are *non-integrated*. This indicates *disalignment* between designers, in particular, a shift between activities such as transitions from one focus to another type of focus: (1) from one focus to another in the problem/solution space (i.e., shift of problem concerned by the activities), or (2) from a focus in one space to a focus in another space (problem/solution space, group space, domain space). In this case, there is, by definition, neither redundancy nor complementarity between modalities.

## 4. Exploiting different modalities for alignment and disalignment

Collaboration between partners in a design meeting can take many different forms. This section presents examples of different ways in which design partners may articulate their activities, exploiting the possibilities provided by different modalities that are available for interaction, i.e. verbal, graphical and gestural, in our current analysis.

### 4.1. Integrated activities

Integrated activities have been identified both in individually and in collectively performed actions. Indeed a designer talking, and drawing or gesturing at the same moment, has been observed frequently during the meeting analysed. In example 1 (see Table 2), Louis is simultaneously generating a solution [GEN/SOLa_PB1] and indicating, with his hand, a two-dimensional zone on the drawing, in order to reinforce his proposal.

**Table 2.** Example of an integrated multimodal design activity (between 12 :08 :27 and 12 :08 :28): one designer (Example 1)

| Line identif | Actor Verb | Actor Gr-Ge | Verbal action (transcription) | Gr-Ge action | Attr1 (obj1) | Attr3 (tool) |
|---|---|---|---|---|---|---|
| 1 | L | L | We reverse the problem and, finally, we do u:h | Delimit_2d | C16+P1 | hand |

Integrated activities have also been observed in collectively performed actions. An example of designers collaborating through integrated activities, using complementary action modalities, concerns the co-elaboration of a solution to a same problem: a same type of activity on the same object (the same problem), implemented through graphical, gestural and verbal actions (see Example 2 in Table 3).

**Table 3.** Example of an integrated multimodal design activity (between 12:08:28 and 12:08:38): two designers co-elaborating a solution, through graphical, gestural, and verbal actions (Example 2)

| Line identif | Actor Verb | Actor Gr-Ge | Verbal action (transcription) | Gr-Ge action | Attr1 (obj1) | Attr2 (obj2) | Attr3 (tool) |
|---|---|---|---|---|---|---|---|
| 1 | | L | | Graph_trac | C16 | C16_over_P1 | pencil |
| 2 | | L | | Delimit_2d | C16+P1 | | hand |
| 3 | C | | on both sides here reducing (.) but there there what is a pity is that one has a beautiful vaulted hall | | | | |
| 4 | | C | | Graph_trac | C16 | C16_over_P1 | |

Louis who, previously, has formulated a solution proposal starts to sketch it (line 1), underlining his proposal by a hand gesture (line 2); Charles continues, detailing the solution, consecutively in a verbal (line 3) and a graphical (line 4) mode. Louis' gestural underlining of his proposal is redundant with his graphical elaboration. Charles' detailing of the proposal is complementary with Louis' elaboration. All these segments are coded as solution-generation activities [GEN/SOLa_PB1]. One may notice that they are expressed through various modalities.

### 4.2. Non-Integrated activities

Our last example is more complex. It shows both integrated (INT) and non-integrated (Non-INT) activities. It is composed of three parts (INT, Non-INT and INT), whose global structure is INT—Non-INT//INT, that is, (1) an integrated activity is (2) interrupted and followed by a non-integrated activity, following which (3) an integrated and a non-integrated activity continue in parallel (see Example 3 in Table 4).

All three designers are involved. An integrated activity by two designers (Marie and Charles, line 1 to 4) is interrupted by the third one (Louis, line 5) who attacks another non-integrated activity, which he pursues in parallel with his two design colleagues who continue, now in coalition, their joint solution elaboration (line 6 to 14).

Marie's verbal formulation and evaluation of a solution proposal for the bar in the little lounge ([GEN/SOLa_PB1] and [EVAL+/SOLa_PB1], lines 1 and 3), overlaps with a positive evaluation that Charles formulates concerning her proposal ([EVAL+/SOLa_PB1], line 2). Marie's overlapping turn is a collaborative one and she simply continues her proposal, gesturally "drawing" the proposal with her pencil. Then Louis interrupts the bar-elaboration activity by the verbal proposal of a solution for another problem (the lift) ([GEN/SOLb_PB2], line 5). Within less than a second, Marie continues her elaboration of the bar ([GEN/SOLa_PB1], line 6), so that Louis and she, during a split second, come to work in parallel. Marie carries on, while Louis breaks of —at least, the explicit expression of his activity. Marie continues to elaborate the bar solution ([GEN/SOLa_PB1], lines 7 and 10). This continuation of the bar elaboration is both an individually integrated activity (Marie elaborating her solution idea verbally and graphically) and a collectively integrated activity (Marie's solution elaboration being supported by Charles, lines 8 and 12). However, Louis comes back, and the parallel non-integrated activity on the bar (PB1: M and C) and the lift (PB2: L), which had started at line 5, continues during another eight turns.

**Table 4.** Example of a composite [integrated – non-integrated // integrated] multimodal design activity (between 12:07:51 and 12:08:38): a coalition of two designers (M and C) co-elaborating a solution for one problem, and a third one (L) elaborating a solution for another problem (Example 3)

| Line identif | Actor Verb | Actor Gr-Ge | Verbal action (transcription) | Gr-Ge action | Attr1 (obj1) | Attr2 (obj2) | Attr3 (tool) |
|---|---|---|---|---|---|---|---|
| 1 | M | | that that would have been a space | | | | |
| 2 | C | | yes better= | | | | |
| 3 | M | | =ideal for the bar and there yes when one is there one feels that: the- there something is taking place that will be:= | | | | |
| 4 | | M | | Movem_2d | C16+P1 | | pencil |
| 5 | L | | =or indeed if one decides to [di- to dig | | | | |
| 6 | M | | [xxx little sounds or: while there there it is it is still \ and it is too far from there to go and sit there to wait | | | | |
| 7 | | M | | Point | C16+P1 | | pencil |
| 8 | C | | yes | | | | |
| 9 | | | (..) | | | | |
| 10 | M | M | it:is (:) in fact one is waiting over here | Point | C16+P1 | | hand |
| 11 | | | (…) | | | | |
| 12 | C | | yes (..) that's true | | | | |
| 13 | L | | no indeed if one decides to dig one coul[::d | | | | |
| 14 | M | L | [or otherwise one may wait in the bar | Position | C_Virgin | C16 | hand |

## 5. Discussion

We have analysed and illustrated two modes of articulation between modalities in parallel interactions, which reflect either alignment or disalignment between co-designers. In the integrated mode, co-designers share their concern with a particular category of object, each designer performing the same or a different action on it, through identical (e.g. verbal) or different (e.g. verbal and graphical) modalities. This reflects an alignment between designers with respect to the sharing of a same focus of work [37]. Furthermore, there is a semantic redundancy or complementarity between the different semiotic modalities (Ex. 1 and 2). In the non-integrated mode, co-designers work on different objects (pertaining to different categories) and pursue different local goals. This translates a gap between the designers' focus, resulting for example, from a shift from one design problem to another one (Ex 3). At a more general level, this also indicates shifts between spaces of representation (problem/solution space, group space, domain space).

It is worth to discuss the present analysis and results with respect to the coalition concept. Indeed, our analysis, based on a triadic design situation, illustrates a coalition process: alignment of two co-designers with respect to a problem, combined with opposition towards a third co-designer focused on another problem. This coalition is of another nature than the ones analysed in pragmatic-linguistics analyses of trilogs based on verbal corpora [38]. These linguistics studies show that relationships between participants in a meeting between three people sharing the same global focus can be of various natures, particularly convergence versus divergence with respect to theses or proposals in an argumentation process. As soon as three people are together, coalitions between two of them against the third one may appear. Caplow [1971, quoted in 39, p. 54] even defends that it is one of the essential characteristics of triadic conversations. Zamouri [39] concludes, based on the analysis of a verbal corpus, that coalitions always emerge from a conflict, which may be initiated, for example, by a counter-proposal.

Such a kind of coalition linked to the argumentation process could be involved in our integrated mode, were co-designers are aligned with respect to the same problem-focus and develop alternative solutions concerning which they may agree or not. For example, there could be a coalition of two designers —one generating verbally a solution, while the second draws it— against a third designer generating an alternative solution, be it verbally or graphically.

An original contribution of our study is to show that coalitions may also occur at another level, with respect to gaps in the focus of people's work. Our third example illustrates this kind of coalition: two designers working on one problem while the third one works on another problem. One could also have coalitions between designers working in different categories of spaces (problem/solution space, group space, domain space).

Another original contribution of the present study is to show that coalitions may be expressed not only in a verbal mode, but also through particular articulations of different semiotic systems ("modalities").

An important issue will be to understand whether coalitions and disalignments with respect to categories of spaces may be disturbing or on the contrary may help designers to advance in their work. Still another issue is whether these kinds of coalition reflect disagreements between designers. Although further work is necessary to handle these questions, we can already advance some reflections based on different cases of disalignments.

- Disalignments between problem space and group space: One or two designers deal with a problem/solution in the problem space while one or two others start setting up another goal or another task in the group space.
    - o This case may reflect an implicit disagreement on the completeness of the solution at hand. Some of the designers (but not all) consider the solution is complete and try to skip to another task.
    - o This case may also indicate that some of the designers (but not all) evaluate that the problem at hand has some relationships with another task, which then can be interesting to deal with at this point.
- Disalignments between problem space and domain space: One or two designers deal with a problem/solution in the problem space, while one or two others switch to exchanges on domain objects, domain rules or procedures.
    - o Again, this case may reflect an implicit disagreement on the solution at hand. However, the disagreement is not on the completeness but rather on the adequacy of the solution. Some of the designers (but not all) consider the solution is inadequate and refer to domain knowledge that is relevant for an argumentation move.
    - o This case may also reflect a thematic drift, triggered by the problem/solution at hand. Whereas this drift does not provide knowledge required for evaluating the solution at hand, it is useful in a cognitive synchronization process.

Further work could examine these different cases based on protocol data and could search for other cases of disalignment. To this end, we believe that the methodological framework that we have developed can offer great potentialities to systematise the identification and statistical analysis of various types of coalitions and disalignments: search for patterns of co-occurrence of both graphico-gestural and verbal activities with respect to different spaces; search for combinations of these co-occurrences with sequential patterns. In this objective, we believe that the methodological cost of both developing our coding scheme and applying it to a corpus is compensated by the possibilities of treatments they offer, in contrast with a more narrative approach.

With respect to the development of computer tools, accounting for parallel activities is of particular interest for the support of synchronous collaborative design. Our results show that simultaneous activities may occur "naturally" through various modalities and that the forms of articulation between modalities are meaningful with respect to people's alignment regarding their work. Coming to understand the way in which people are aligned or not concerning their focus of work is very important for such kind of devices. Our work is preliminary regarding this issue.

Further work could examine the way in which alignments and disalignments are expressed through particular shifts between modalities. We did not analyse the role of attention in this text. Different forms of articulation require more or less attention on behalf of the partners who are interacting. Two partners in alignment may exploit the same modality at the same moment, but, if cooperative design elaboration is aimed, such a form of articulation will generally lead to problems of attention. Using different modalities can thus offer ways to progress together without disturbing attention. In disalignment situations, special attention from others might be required. Using the same modality (as in Example 3) can be a way to engage others to shift focus.

**Acknowledgements**

This research was funded by the COGNITIQUE program "Cognitions, interactions sociales, modélisations" (MOSAIC project).